\documentstyle[twocolumn]{mn}
\input{psfig.sty}

\title[Polarization of radio cores and the unified schemes]{Polarization of
radio cores as a test of the unified schemes} 

\author[D.J. Saikia ]{D.J. Saikia \thanks{E-mail: djs@ncra.tifr.res.in, djs@gmrt.ernet.in} \\
National Centre for Radio Astrophysics, TIFR,
Post Bag 3, Ganeshkhind, Pune 411 007, India \\
}

\date{Received:}
\begin{document}

\maketitle

\begin{abstract}
In the unified schemes for extragalactic radio sources, the quasars and
BL Lac objects are intrinsically similar to the Fanaroff-Riley class II and I
radio galaxies respectively, but appear to be different because they are 
viewed at different angles to the line-of-sight. The quasars and BL Lac
objects are observed at small viewing angles while the radio galaxies lie
close to the plane of the sky. In such schemes, the quasar or BL Lac nucleus
is hidden from our view by a putative torus when the source is inclined at
a large angle to the line-of-sight and appears as a radio galaxy. 
Such  a scenario should also 
affect the observed polarization properties of the cores, with the cores
in the radio galaxies and weak-cored quasars being less polarized than the
core-dominated quasars and BL Lacs due to Faraday depolarization by the
magnetoionic medium in the obscuring torus or disk. In this paper,
we report that the core polarization of radio galaxies and weak-cored 
quasars is indeed much smaller than for the core-dominated
quasars and BL Lacs. We suggest that this is
due to the depolarization caused by the obscuring torus or disk, providing
further evidence in favour of the unified schemes. 

\end{abstract}

\begin{keywords}
galaxies: active - galaxies: nuclei - quasars: general - BL Lacertae objects: 
general - radio continuum: galaxies - polarization 
\end{keywords}

\section{Introduction}

The polarization properties of radio cores in active galactic nuclei
could provide valuable information on the physical conditions in the
nuclear regions of active galaxies (Rudnick \& Jones 1983; O'Dea 1989;
Udomprasert et al. 1997; Saikia et al. 1998), and also provide tests of the
unified scheme for powerful radio sources (cf. Saikia 1995; Saikia \& 
Kulkarni 1998, hereinafter referred to as SK98). In the unified 
scheme for Fanaroff \& Riley (1974) class II or FRII radio sources, 
the radio galaxies and quasars are intrinsically similar objects but 
appear to be different because they are inclined at different angles to the
line-of-sight. The radio galaxies lie
close to the plane of the sky while their relativistically 
beamed counterparts, the quasars, lie within about 45$^\circ$ to
the line-of-sight. In an equivalent scheme for the lower-luminosity 
FRI radio galaxies, the relativistically beamed objects seen at small
angles to the line-of-sight are the BL Lac objects.
(Scheuer 1987; Peacock 1987;
Barthel 1989; Browne \& Jackson 1992; Antonucci 1993; Urry \& Padovani 1995). 
In these unified schemes 
the quasar or BL Lac nucleus is hidden from our view by a putative torus when the
source is inclined at a large angle to the line-of-sight and is observed
as a radio galaxy. The radio core observed through such a torus or disk
might exhibit weaker levels of polarization and higher values of rotation measure (RM) compared
with the quasars or BL Lac objects where we have a direct view of the nucleus or radio
core (cf. Saikia 1995). 

The obscuring torus or disk is unlikely
to have a sharp boundary and one might find a lower degree of polarization
and higher RM in the cores of lobe-dominated quasars as they are seen through
the edge of the disk or torus. SK98
have examined recently the core polarization properties
of core- and lobe-dominated quasars and have shown that the degree of core polarization
of the lobe-dominated quasars is indeed lower, consistent with the ideas of the
unified scheme.  They have also examined the relative orientation of the core-polarization 
E-vector at $\lambda$6cm and the radio axis for the weak-cored quasars. The sample is
small but does not show a significant trend reported earlier for cores of
moderate strength (Saikia \& Shastri 1984). SK98 suggest that this could also be due to 
Faraday rotation by the material in the edge of the torus or disk. 

In this paper we extend the study of SK98, who confined themselves to quasars, to
include both FRI and II radio galaxies and BL Lac objects. We compare the degree of core 
polarization, p$_c$,  of the two FR types of radio galaxies with the 
core polarization of quasars and
BL Lacs and examine whether these are consistent with the basic ideas of the two
equivalent unified schemes.

\begin{table}
{\bf Table 1.} The sample of sources \\
\vspace*{1ex}
\begin{tabular}{l   l l l r l }
Source     & Id. &  Red-  &  p$_c$   &   f$_c$    & Refs. \\
           &     &  shift &  \%      &        &        \\
\\[1mm]
0040+517  &   G & 0.174   & $<$1.70       &  0.0012    &         22  \\
0048$-$097  &   B & 1     & 3.20$\pm$0.20 &  0.52      &    1,2,3  \\
0055+300  &   G & 0.0167  & $<$0.30       &  0.34      &   2,6,7,8  \\
0104+321  &   G & 0.0167  & 1.00$\pm$0.50 &  0.083     &       3,8  \\
0109+224  &   B & 1       & 9.40$\pm$0.30 &  0.88      &    1,2,3  \\
0118$-$272  &   B & 0.557  & 6.50$\pm$0.30 &  0.69      &   1,2,3  \\
0138$-$097  &   B & 0.501  & 4.10$\pm$0.20 &  1         &    1,2,3  \\
0220+427  &   G & 0.0215  & 0.50$\pm$0.01 &  0.12      &     9,10  \\
0235+164  &   B & 0.940   & 0.40$\pm$0.10 &  0.96      &    1,2,4  \\
0305+039  &   G & 0.0289  & $<$0.60       &  0.27      &   2,3,12  \\
0307+169  &   G & 0.2559  & $<$0.40       &  0.0066    &         22  \\
0314+416  &   G & 0.0255  & $<$0.80       &  0.011     &      3,11  \\
0316+413  &   G & 0.0172  & 0.12$\pm$0.04 &  1         &    2  \\
0326+396  &   G & 0.0243  & $<$0.40       &  0.25      &       13 \\ 
0338$-$214  &   B & 1       & 2.10$\pm$0.20 &  0.96      &    1,2,3  \\
0406+121  &   B & 1.02    & $<$0.30       &  0.76      &    1,2,3  \\
0415+379  &   G & 0.0485  & 0.95$\pm$0.03 &  0.28      &      14  \\
0422+004  &   B & 1       & 8.40$\pm$0.20 &  1         &    1,2,3  \\
0430+052  &   G & 0.033   & 4.86$\pm$0.07 &  0.51      &   2,3,15  \\
0433+295  &   G & 0.2177  & $<$0.20       &  0.0090    &         22  \\
0453+227  &   G & 0.214   & $<$0.90       &  0.0048    &        22  \\
0454+844  &   B & 0.112   & 3.70$\pm$0.20 &  0.86      &    1,2,3  \\
0512+249  &   G & 0.064   & $<$2.90       &  0.0014    &      23,25  \\
0537$-$441  &   B & 0.896   & 1.00$\pm$0.10 &  1         &    1,2,3  \\
0651+542  &   G & 0.2384  & $<$2.90       &  0.0023    &         22  \\
0716+714  &   B & 1       & 2.80$\pm$0.40 &  0.46      &    1,2,3  \\
0734+805  &   G & 0.1182  & $<$1.00       &  0.0066    &      23,25  \\
0735+178  &   B & 0.424  & 1.00$\pm$0.10 &  1         &    1,2,3  \\
0808+019  &   B & 1       & $<$0.80       &  0.32      &    1,2,5  \\
0814+425  &   B & 1       & 0.50$\pm$0.10 &  0.87      &    1,2,4  \\
0818+472  &   G & 0.1301  & $<$0.90       &  0.016     &     24,25  \\
0823+033  &   B & 0.506   & 2.20$\pm$0.20 &  0.81      &    1,2,3  \\
0823$-$223  &   B & 0.910   & 3.60$\pm$0.10 &  1         &    1,2,5  \\
0828+493  &   B & 0.548   & 1.80$\pm$0.20 &  0.92      &   1,2,3  \\
0851+202  &   B & 0.306   & 4.50$\pm$0.10 &  0.90      &    1,2,3  \\
0917+458  &   G & 0.1745  & 0.41$\pm$0.08 &  0.030     &      3,16  \\
0938+399  &   G & 0.1075  & $<$1.70       &  0.010     &     24,25  \\
0945+076  &   G & 0.0861  & $<$1.10       &  0.0057    &      24,25  \\
0954+658  &   B & 0.367   & 4.40$\pm$0.30 &  0.48      &    1,2,3  \\
1144$-$379  &   B & 1.048   & 1.60$\pm$0.10 &  1         &    1,2,5  \\
1147+245  &   B & 1       & 1.10$\pm$0.30 &  0.79      &    1,2,4  \\
1219+285  &   B & 0.102   & 2.20$\pm$0.10 &  1         &    1,2,5  \\
1221+809  &   B & 1       & 4.30$\pm$0.50 &  0.84      &    1,2,3  \\
1308+277  &   G & 0.2394  & $<$1.70       &  0.0066    &         22  \\
1322$-$427  &   G & 0.002   & 0.21$\pm$0.01 &  0.017     &     17,26  \\
1413+135  &   B & 0.247   & $<$0.20       &  1         &    1,2,5  \\
1418+546  &   B & 0.152   & 2.60$\pm$0.10 &  1         &    1,2,5  \\
1514$-$241  &   B & 0.048   & 7.40$\pm$0.10 &  1         &    1,2,5  \\
1519$-$273  &   B & 1       & 2.60$\pm$0.10 &  0.97      &    1,2,3  \\
1538+149  &   B & 0.605   & 8.80$\pm$0.10 &  0.85      &    1,2,4  \\
1637+826  &   G & 0.023   & $<$0.30       &  0.56      &   2,18,21  \\
1717+178  &   B & 1       & 5.60$\pm$0.30 &  0.69      &    1,2,4  \\
1749+701  &   B & 0.770   & $<$0.40       &  0.65      &    1,2,3  \\
1803+784  &   B & 0.680   & 5.20$\pm$0.10 &  0.79      &      1,2  \\
1807+698  &   B & 0.050   & 2.70$\pm$0.10 &  0.82      &    1,2,4  \\
1823+568  &   B & 0.663   & 5.60$\pm$0.20 &  0.79      &   1,2,4  \\
1832+474  &   G & 0.1605  & $<$1.30       &  0.0042    &         22  \\
1833+326  &   G & 0.0578  & $<$0.10       &  0.18      &    24,25  \\
1842+455  &   G & 0.0917  & $<$0.70       &  0.051     &      3,19  \\
1845+797  &   G & 0.0569  & $<$0.60       &  0.11      &     3,20  \\
1921$-$293  &   B & 0.352   & 3.40$\pm$0.10 &  0.74      &      1,2  \\

\end{tabular}
\end{table}

\begin{table}
{\bf Table 1.} Contd.
\vspace*{1ex}

\begin{tabular}{l   l l l r l }
Source     & Id. &  Red-  &  p$_c$   &   f$_c$    & Refs. \\
	   &     &  shift &  \%      &        &        \\
\\[1mm]
1939+605  &   G & 0.201   & $<$0.10       &  0.027     &        22  \\
2007+777  &   B & 0.342   & 4.20$\pm$0.10 &  1         &    1,2,5  \\
2010+723  &   B & 1       & 5.40$\pm$0.20 &  0.94      &      1,2  \\
2045+068  &   G & 0.1270  & $<$1.20       &  0.017     &     24,25  \\
2121+248  &   G & 0.1016  & $<$3.80       &  0.00050   &      24,25  \\
2131$-$021  &   B & 1.285   & 2.80$\pm$0.10 &  1         &    1,2,5  \\
2150+173  &   B & 1       & 0.90$\pm$0.30 &  0.62      &      1,2  \\
2153+377  &   G & 0.290   & $<$0.20       &  0.015     &        22  \\
2200+420  &   B & 0.069   & 0.80$\pm$0.10 &  1         &    1,2,5  \\
2254+074  &   B & 0.190   & 1.60$\pm$0.50 &  0.75      &    1,2,5  \\

\end{tabular}

\vspace*{1ex}
References 1: V\'{e}ron-Cetty  \& V\'{e}ron 1998; 2: Perley 1982; 3: K\"{u}hr
et al. 1979; 4: K\"{u}hr et al. 1981; 5: NASA Extragalactic Database; 6: Mack
et al. 1997; 7: Venturi et al. 1993; 8: Fomalont et al. 1980; 9: Hardcastle
et al. 1996; 10: Leahy, J\"{a}gers \& Pooley 1986; 11: O'Dea \& Owen 1986;
12: Saikia et al. 1986; 13: Bridle et al. 1991; 14: Linfield \& Perley 1984;
15: Walker, Benson \& Unwin 1987; 16: Clarke et al. 1992; 17: Burns, Feigelson \& Schreier 1983;  
18: Perley, Bridle \& Willis 1984; 19: Burns, Christiansen \& Hough 1982;
20: Dreher 1981; 21: Saripalli et al. 1986; 22: Hardcastle et al. 1997; 
23: Leahy et al. 1997; 24: Black et al. 1992; 25: Hardcastle et al. 1998;
26: Junkes et al. 1993

\end{table}

\section{The sample of sources}

In this paper we have considered both FR class I and II radio galaxies and a sample of BL Lac
objects for which either measured  values or good upper limits to their degree of core polarization 
at about 5 or 8 GHz are available or could be estimated  from images with angular resolutions
better than about an arcsec. Observations with high angular resolution are required to  
minimize contamination from the extended emission which could be quite strongly polarized.
This is particularly important for the FR class I sources where the radio jets are
often quite prominent close to the radio cores and are contiguous with it. 

\begin{figure*}
\vbox{
\vspace{-6.0in}
\psfig{figure=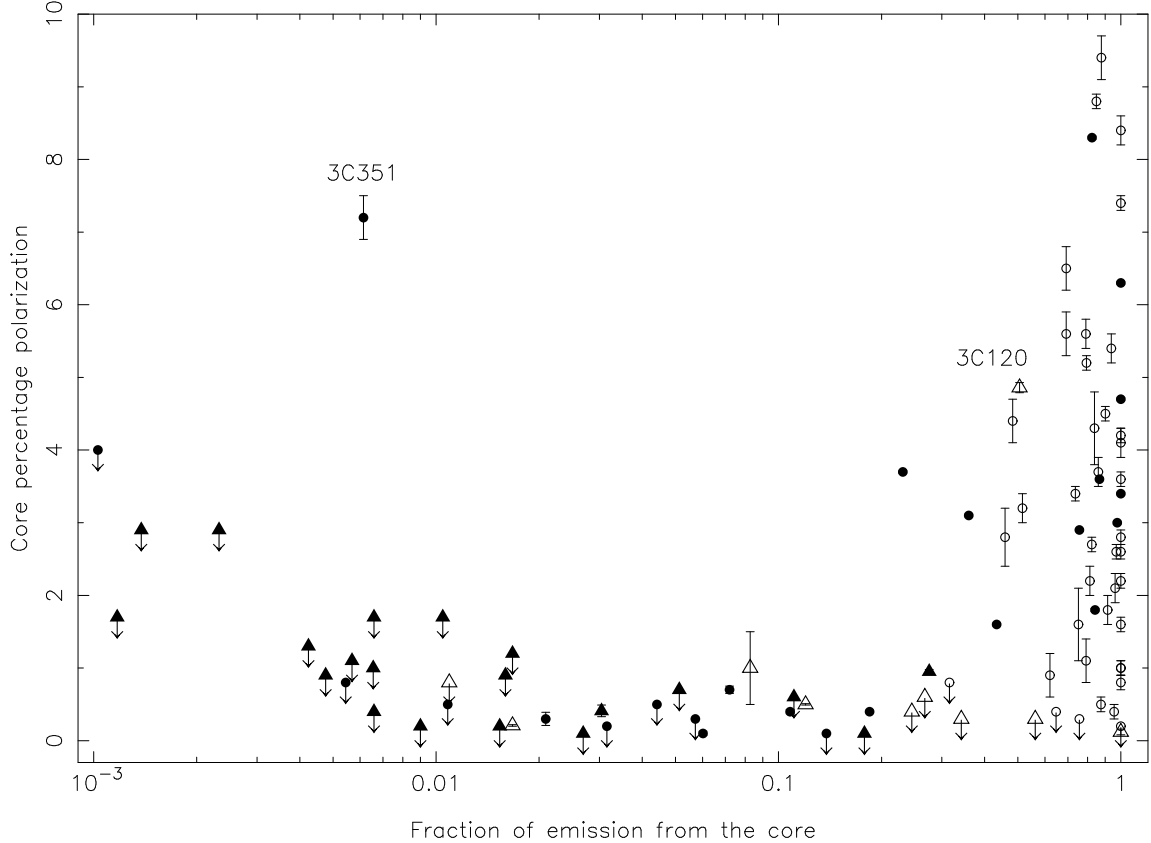}
\vspace{-0.0in}
\caption{ The fraction of emission from the core at an emitted frequency of 8 GHz plotted
against the percentage polarization of the radio core at $\lambda$6 or 3.6cm for BL Lacs
(open circles), quasars (filled circles), FRI radio galaxies (open triangles) and FRII
radio galaxies (filled triangles). The arrow indicates an upper limit. The two sources
3C120 and 3C351 discussed in the text are indicated. }
}
\end{figure*}

The sample of sources has been compiled from the literature and is listed in Table 1, which is arranged
as follows. Column 1: source name in the IAU convention; column 2: optical identification
where B and G denote a BL Lac object and radio galaxy respectively; column 3:
redshift where we have adopted a value of 1 for sources without a measured redshift;
column 4: polarization percentage of the radio core, p$_c$,
at 5 or 8 GHz and its error; 
column 5: the fraction of emission from the core, f$_c$,  at an emitted frequency of 8 GHz;
column 6: references for 
the degree of core polarization and the core and total flux density of the objects which have
been used for estimating f$_c$.
The fraction of emission from the core, f$_c$, is used as a statistical indicator of orientation
of the source axis to the line of sight, and has been estimated using a spectral index, $\alpha$
defined as S $\propto \nu^{-\alpha}$, of 0 for the core and 1 for the extended emisssion.
If the measured flux density of the core in the high-resolution observations for determining the 
degree of core polarization is larger than the listed values of the total flux density of the source in
both K\"{u}hr et al. (1979, 1981) and the NASA Extragalactic Database, then the 
total flux density has been set equal to the  core flux density.
For the FR II radio galaxies which have been observed by the Cambridge group with 
high resolution and sensitivity using the VLA at about 8 GHz (Black et al. 1992; Leahy et al. 1997;
Hardcastle et al. 1997, 1998), we have estimated the upper limits using their listed values of
the core flux density and 3 times the rms noise in the polarization maps for
all the cores where it could be ascertained from their images that there is no core
polarization. For the case of 3C123 (0433+295) the upper limit has been taken to be 5 times the rms noise.
The sample of BL Lacs consists of all the objects from the recent list of V\'{e}ron-Cetty
\& V\'{e}ron (1998) which have been observed by Perley (1982) with the VLA A-array
at 5 GHz. The error in the fractional polarization has been estimated using the expression
given by Perley (1982) for the error in the polarized flux density.

\section{Results and discussion}

In Figure 1, we plot the fraction of emission from the core, f$_c$
against the degree of polarization from the core for the FR I and II radio galaxies and
BL Lacs listed in Table 1. In addition, we have also plotted the sample of lobe- and
core-dominated quasars listed by SK98. In SK98, the polarization of the
weak cores in the lobe-dominated quasars are from the high-sensitivity and
high-resolution observations of  Bridle et al. (1994).
The sample of core-dominated quasars have been observed with the
VLA A-array at $\lambda$6\,cm by Saikia et al. (1998). The median values and the
range of the degree of core polarization for this sample of  core-dominated quasars 
is similar to those measured by O'Dea (1989) but marginally higher than for a large
sample of VLA calibrators where the median value is about 2 per cent 
(cf. Perley 1982; Saikia, Singal \& Cornwell 1987). The sample of Saikia et al. (1998)
was chosen to determine the RMs of cores, and their slightly higher value is due
to this selection effect.

It is clear from Figure 1 that the core-dominated quasars and the BL Lac objects
which are believed to be inclined at small angles to the line-of-sight in the
unified schemes are similarly polarized with a median value of about 3 per cent.
The degree of core polarization drops for sources with
weaker cores. The median value of p$_{c}$
for the weak-cored quasars, viz those with f$_c \leq$ 0.2, 
is less than about 0.4 per cent. The median values for the FR class I and II radio 
galaxies are less than about 0.5 and 0.9 per cent respectively. These median values
are upper limits since no polarization from the core has been detected for most of
the sources. Since most of the galaxies  and weak-cored
quasars have upper limits to p$_{c}$, and the values of core polarization for
some of the FRI radio galaxies could have been affected by prominent jets close
to the nucleus, it is difficult to establish from the present data
whether the cores of radio galaxies are less polarized than those of
the weak-cored quasars. However, the trends for the cores of FRI radio galaxies 
to be less polarized than those of the BL Lacs, and 
the cores of FR II galaxies and weak-cored quasars to be less polarized
than the core-dominated quasars are consistent with the unified schemes.
A Kolmogorov-Smirnoff test shows the distributions of core polarization 
for the above cases to be different at a significance of level of over 99.99 per cent.
It is worth mentioning that the VLBI-scale cores in galaxies also tend to be 
less polarized than those in quasars and BL Lac objects (cf. Cawthorne et al. 1993a, b).

Our results are unlikely to be due differences in redshift between the different 
kinds of objects. The core- and lobe-dominated quasars have a similar median redshift of
about 0.75 and yet exhibit a striking difference in the degree of core polarization. 
Again, considering only those BL Lacs with a redshift less than 0.2, which is
the case for most of the galaxies, the median value of p$_c$ for the BL Lacs is 
again $\sim$ 2.6 per cent, similar to that of the entire sample. 
It is also relevant to note that in
core-dominated sources with multifrequency polarization observations, there is little
evidence of any significant dependence of core polarizaton on wavelength,
possibly  due to several complicating
factors. These are primarily wavelength-dependent opacity effects with different
electron populations contributing at different wavelengths and also variability at
different wavelengths (cf. Saikia et al. 1998). Thus although the integrated 
core polarization of weak-cored quasars is expected to be lower, they might not 
exhibit a smooth p-$\lambda$ relationship (cf. SK98).

Two objects in our sample which merit particular attention are the weak-cored quasar
3C351 (1704+608) which has a core polarization of 7.2 per cent, and the radio galaxy 
3C120 (0430+052) which is
nearly 5 per cent polarized. In 3C351, the nucleus appears extended with an angular
resolution of about 1 arcsec but is resolved into two knots of similar brightness, 
one of which is the core, and the other a knot in a 
prominent jet whose polarization ranges from 11 to 31 per cent (Bridle et al. 1994).
The high polarization is possibly due to some contribution from this prominent jet close
to the nucleus. The radio galaxy 3C120 which has a strongly polarized core
has the signatures of a source inclined close to the line-of-sight although it is a radio galaxy.
Based on its emission-line properties, it has been classified as either a broad line
radio galaxy or a Seyfert galaxy although its spiral structure is not clearly established
(cf. Zirbel \& Baum 1998). It exhibits superluminal motion (cf. Walker, Benson
\& Unwin 1987) and the inferred maximum angle of orientation of the source axis (Cohen 1990)
is within the approximate dividing line for radio galaxies and quasars in the unified scheme
(Barthel 1989). The nucleus is strong and variable at all wavelengths from radio to X-ray
wavelengths (Maraschi et al. 1991) and in many ways it resembles a miniquasar.

The lower degree of core polarization for galaxies and weak-cored quasars  would be expected 
in the unified scheme if their nuclear or core  emission 
are seen through a magnetoionic medium
which forms a part of the obscuring torus or disk. There is likely to be a significant ionized
component in the disk or torus, in addition to the molecular and
atomic components, due to X-ray or UV heating from the active nucleus (cf. Levinson,
Laor \& Vermeulen 1995).
Some evidence for ionized gas in a disk comes from the observations of the FRI radio galaxy 3C84
(0316+413) associated with NGC1275, and the compact symmetric object 4C31.04. In 3C84 
VLBI observations show a prominent jet towards
the south with evidence of subluminal motion, and a weak
counterjet. The spectra of the counterjet show that it has
an abrupt cutoff at low frequencies which is not consistent
with synchrotron self absorption but could be understood
in terms of free-free absorption by a disk which has been
ionized by the central continuum source. The disk affects
the spectrum of the counter jet which is seen through the
disk without affecting the spectrum of the prominent jet
(Vermeulen, Readhead \& Backer 1994; Walker, Romney \&  Benson 1994;
Levinson et al.  1995; Silver, Taylor \& Vermeulen 1998). In the 
compact symmetric source 4C31.04 there is also evidence of free-free absorption due to
individual clouds which are evaporating off the inner edge of the disk (Conway 1996). 
Assuming that the depolarization of the cores in galaxies and weak-cored quasars are  being caused by
a Faraday screen where the RM is a Gaussian random field with dispersion $\sigma$,
the values of  $\sigma$ which are required to decrease the observed polarization from
about 2.5  to 0.4 per cent  are in the range of 250 to 500 rad m$^{-2}$ when the 
ratio of the scale of the RM fluctuations to the resolution varies from 0 to 1
(Tribble 1991, 1992; SK98).

\section{Concluding remarks}
We report the degree of polarization of the radio cores at 5 or 8 GHz  to be
significantly smaller for both Fanaroff-Riley class I and II radio galaxies and 
weak-cored quasars compared to the core-dominated quasars and BL Lac objects. The 
present sensitivity is not adequate to determine any difference between the weak-cored quasars
and radio galaxies. We suggest that the lower degree of core polarization for galaxies
and weak-cored quasars is due to Faraday dispersion by the magnetoionic component in 
a putative torus or disk which obscures a direct view of the nucleus for both FRI and II radio galaxies.
In this scenario, the low values for the weak-cored quasars are due to Faraday dispersion by
the edge of  the obscuring torus. The higher degree of core polarization for the core-dominated
quasars and BL Lacs is due to their smaller angles of inclination to the line-of-sight so that they
are not obscured by the disk or torus. These trends provide additional evidence in favour of the
two equivalent unified schemes, one  for the FRI radio galaxies and BL Lac objects, and the other
for the FRII radio galaxies and quasars.

\section*{ACKNOWLEDGMENTS}
It is a pleasure to thank S. Jeyakumar for his help, 
Judith Irwin, C.H. Ishwara-chandra, Vasant Kulkarni and an anonymous referee for their 
helpful comments on the manuscript, and my colleagues at NCRA for several useful discussions.
This research has made use of the NASA/IPAC extragalactic database (NED)
which is operated by the Jet Propulsion Laboratory, Caltech, under contract
with the National Aeronautics and Space Administration.

%\vspace{0.5cm}

\end{document}